\documentclass[reprint,aps,pra,amsmath,amssymb,groupedaddress]{revtex4-2}
\usepackage{graphicx}
\usepackage{dcolumn}
\usepackage{bm}
\usepackage{color} 
\usepackage[tight]{subfigure}
\usepackage{hyperref}

\newcommand{\ket}[1]{\left|{}#1 \right>}
\newcommand{\bra}[1]{\left<{}#1 \right|}
\newcommand{\expect}[1]{\left<{}#1\right>}
\newcommand{\interproduct}[2]{\langle {}#1 | {}#2 \rangle}

\newcommand{\tr}[2]{\mathrm{tr}_{{}#2} \left[{}#1\right]}

\newcommand{\rmi}{\mathrm{i}}
\newcommand{\rme}{\mathrm{e}}

\begin{document}
	
	\title{Quantum walk for SU(1,1)}
	\author{Liwei Duan}
	\email{duanlw@gmail.com}
	\affiliation{Department of Physics, Zhejiang Normal University, Jinhua 321004, China}
	
	\date{\today}
	
	\begin{abstract}
		We propose a scheme to implement the quantum walk for SU(1,1) in the phase space, which generalizes those associated with the Heisenberg-Weyl group. The movement of the walker described by the SU(1,1) coherent states can be visualized on the hyperboloid or the Poincar\'{e} disk. In both one-mode and two-mode realizations, we introduce the corresponding coin-flip and conditional-shift operators for the SU(1,1) group, whose relations with those for Heisenberg-Weyl group are analyzed. The probability distribution, standard deviation and the von Neumann entropy are employed to describe the walking process. The nonorthogonality of the SU(1,1) coherent states precludes the quantum walk for SU(1,1) from the idealized one. However, the overlap between different SU(1,1) coherent states can be reduced by increasing the Bargmann index $k$, which indicates that the two-mode realization provides more possibilities to simulate the idealized quantum walk.
	\end{abstract}
	
	\maketitle
	
	\section{Introduction}
	
	The classical random walk has been employed as a framework to understand and utilize the stochastic processes in a wide spectrum of scientific disciplines \cite{barber1970random}. The superposition and entanglement inherent in the quantum theory introduce some novel perspectives. The combination of quantum theory with the classical random walk provides a new framework, \textit{i.e.} quantum walks \cite{PhysRevA.48.1687,doi:10.1080/00107151031000110776,Venegas-Andraca2012}. A key difference between quantum and classical random walks is the enhanced rate of spreading: the quantum walk reveals a quadratic increase in the variance due to the quantum interference effects \cite{doi:10.1080/00107151031000110776,Venegas-Andraca2012}. It is also noteworthy that the quantum entanglement is inherent in the quantum walk \cite{Venegas-Andraca2012,Carneiro_2005,PhysRevA.73.042302}, which distinguishes it from its classical counterpart.
	
	Quantum walks can be regarded as one of the most fruitful outcomes of quantum information theory \cite{KADIAN2021100419}, as they underlie the development for other models of computation and help in designing quantum algorithms with speed-ups over the classical ones \cite{doi:10.1142/S0219749903000383,Portugal2013-ng,PhysRevA.58.915}. Quantum walks also pave the way towards the development of universal quantum computation \cite{PhysRevLett.102.180501,doi:10.1126/science.1229957,PhysRevA.81.042330}. In addition to the potential applications in quantum information and quantum computation, quantum walks also serve as a powerful tool to simulate various physical phenomena, such as the Landau-Zener transition \cite{PhysRevLett.94.100602}, Anderson localization \cite{PhysRevLett.106.180403,doi:10.1063/1.3643768,Crespi2013}, dynamic quantum phase transitions \cite{PhysRevLett.122.020501}, quantum-to-classical transition \cite{PhysRevA.65.032310,PhysRevA.67.032304,PhysRevA.67.042305,PhysRevLett.91.130602,doi:10.1126/science.1174436,PhysRevLett.121.070402}, the nontrivial topological phase \cite{PhysRevA.82.033429,Kitagawa2012,Cardano2017,PhysRevLett.118.130501,PhysRevLett.124.050502}, non-Hermitian system \cite{PhysRevA.93.062116,Xiao2017,Xiao2020,PhysRevLett.127.270602}, strongly correlated quantum matter \cite{Preiss2015-oe}, etc.
	Motivated by the rich applications of the quantum walk, various experimental platforms have been proposed or developed, ranging from the trapped ion \cite{PhysRevA.65.032310,PhysRevLett.103.090504,Matjeschk_2012}, NMR \cite{PhysRevA.72.062317}, photonics \cite{PhysRevLett.106.180403,Xiao2017,Xiao2020,PhysRevLett.127.270602}, neutral atoms \cite{doi:10.1126/science.1174436}, Bose-Einstein Condensate \cite{PhysRevLett.121.070402,PhysRevLett.124.050502}, cavity quantum electrodynamics \cite{PhysRevA.67.042305,Xue_2008} to superconducting qubits \cite{doi:10.1126/science.aaw1611,doi:10.1126/science.abg7812}, etc.
	
	Depending on the timing used to perform evolution operators, quantum walks are generally classified into two categories \cite{Venegas-Andraca2012}: discrete and continuous quantum walks. For the discrete quantum walk, the evolution operator of the system is performed only in discrete time steps \cite{PhysRevA.48.1687}, while for the continuous quantum walk, the corresponding evolution operator can be performed with no timing restrictions at all \cite{PhysRevA.58.915}. In this paper, we mainly focus on the former case. The implementation of discrete quantum walks generally requires two basic operations \cite{doi:10.1080/00107151031000110776,Venegas-Andraca2012}: coin-flip operation which determines the state of the coin, and the conditional-shift operation which determines the movement of the walker.
	
	The phase space provides a platform to perform the quantum walk \cite{PhysRevA.65.032310,Omanakuttan_2018,PhysRevLett.103.090504}. 
	Previous studies mainly focus on the phase plane associating with the Heisenberg-Weyl group. Based on the harmonic oscillator, the quantum walks over a circle \cite{PhysRevA.67.042305,Xue_2008,PhysRevLett.118.130501} or a line \cite{PhysRevLett.103.090504,Matjeschk_2012} on the phase plane have attracted persistent attention. 
	The coherent state is feasible in the experiments \cite{RevModPhys.62.867}, which can be chosen to describe the walker. The number of steps of quantum walk is limited by the nonorthogonality of the coherent states \cite{PhysRevA.67.042305,PhysRevLett.103.090504,Matjeschk_2012}. 
	There exist various phase spaces, which are intimately connected with the dynamical group of each physical problem \cite{RevModPhys.62.867}. Recently, the quantum walk in the phase space has been extended to the Bloch sphere, which is closely related to the SU(2) group \cite{PhysRevA.105.042215}. A spin cluster described by the spin coherent state serves as the walker. An additional spin plays the role of a coin, who interacts with the spin cluster and determines its rotation on the Bloch sphere. 
	
	In this paper, we generalize the quantum walk to the SU(1,1) group. The SU(1,1) group has been employed in various branches, such as quantum optics \cite{Wodkiewicz:85,Gerry:01}, Bose-Einstein condensates \cite{PhysRevLett.125.253002,PhysRevLett.125.253401,PhysRevA.102.011301,PhysRevA.104.023307} and quantum metrology \cite{PhysRevA.33.4033,doi:10.1063/5.0004873}, etc. 
	With regard to the quantum walk for SU(1,1), the hyperboloid or the Poincar\'{e} disk provides a platform to visualize the walking process.
	The corresponding condition-shift operator is closely related to that of the quantum walk for Heisenberg-Weyl group, which indicates that one can extend the well-studied experimental setups for Heisenberg-Weyl group to that for SU(1,1). The SU(1,1) coherent state can be used to describe walker's location, whose superpositions can be generated during the walking process and have potential applications in quantum sensing \cite{PhysRevA.103.062405,PhysRevA.106.043704}.
	The paper is structured as follows. In Sec. \ref{sec:HW}, we introduce the Heisenberg-Weyl group, and demonstrate the corresponding quantum walk over a circle on the phase plane. The conditional-shift operator and coin-flip operator are introduced, which will be employed to construct corresponding operators for SU(1,1). In Sec. \ref{sec:SU11}, we first revisit the basic properties of the SU(1,1) group. The  phase space for SU(1,1) corresponds to a hyperboloid or a Poincar\'{e} disk. Then, the quantum walk on the hyperboloid or the Poincar\'{e} disk is proposed for both one-mode and two-mode realizations of the SU(1,1) group. The walker is described by the SU(1,1) coherent states. The probability distribution and the standard deviation are calculated, from which we confirm the quadratically growing variance. The influence of nonorthogonality is also discussed. In Sec. \ref{sec:entanglement}, we introduce the von Neumann entropy to describe the quantum entanglement between the coin and the walker. Finally, a brief summary is given in Sec. \ref{sec:summary}.
	
	\section{Quantum walk related to the Heisenberg-Weyl group} \label{sec:HW}
	
	The quantum walk in phase spaces has been widely studied. 
	We begin by briefly reviewing the quantum walk over a circle on the phase plane \cite{PhysRevA.65.032310,PhysRevLett.118.130501}, which is closely related to the Heisenberg-Weyl group. The phase plane consists of all possible values of position and momentum variables, and the walker's location on the phase plane can be described by the coherent state. The coin-flip and the conditional-shift operators for the Heisenberg-Weyl group are introduced, which will be generalized to those in the SU(1,1) group in Sec. \ref{sec:SU11}.
	
	\begin{figure}[htb]
		\centering
		\includegraphics[scale=0.85]{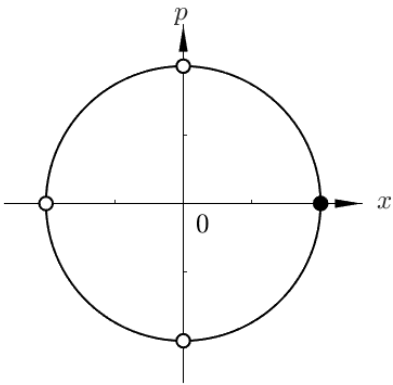} 
		\caption{Schematic diagram of the quantum walk on a circle related to the Heisenberg-Weyl group. The walker located on the phase plane with $\left(x, p\right)$ is described by the HW coherent states $\ket{\alpha}$ with $\alpha = |\alpha| \rme^{\rmi \theta}$. For clarification, $L=4$ sites are shown with $\theta=0$, $\pm \pi / 2$ and $\pi$.}\label{fig:circle}
	\end{figure}
	
	\subsection{Heisenberg-Weyl group}
	
	One usually introduce the bosonic annihilation operator $\hat{a}$ and creation operator $\hat{a}^{\dagger}$ to describe the harmonic oscillator, which satisfy
	\begin{eqnarray}
		\left[\hat{a}, \hat{a}^{\dagger}\right] =  \hat{I}_{\text{a}}, \qquad \left[\hat{a},  \hat{I}_{\text{a}}\right] = \left[\hat{a}^{\dagger},  \hat{I}_{\text{a}}\right] = 0,
	\end{eqnarray}
	with $ \hat{I}_{\text{a}}$ the identity operator. The above commutation relations are described by the Heisenberg-Weyl group, and the corresponding operators can be regarded as the generators of the Heisenberg-Weyl algebra \cite{weyl1928gruppentheorie}.
	
	The coherent state related to the Heisenberg-Weyl group is defined as
	\begin{eqnarray}
		\ket{\alpha} &=& \exp \left(\alpha \hat{a}^{\dagger} - \alpha^* \hat{a}\right) \ket{0} ,
	\end{eqnarray}
	with $\alpha = |\alpha| \rme^{\rmi \theta}$. For clarity, we just call $\ket{\alpha}$ the HW coherent state. 
	The position and momentum operators can be defined as
	\begin{eqnarray}
		\hat{x} = \frac{\hat{a}^{\dagger} + \hat{a}}{\sqrt{2}} , \qquad \hat{p} = \rmi \frac{\hat{a}^{\dagger} - \hat{a}}{\sqrt{2}} .
	\end{eqnarray}
	The harmonic oscillator described by $\ket{\alpha}$ is centered at 
	\begin{eqnarray}
		\left(x, p\right) = \sqrt{2} |\alpha| \left(\cos \theta, \sin \theta\right) ,
	\end{eqnarray}
	with $A = \bra{\alpha} \hat{A} \ket{\alpha}$ ($A=x,p$). As long as $|\alpha|$ is fixed, $(x,y)$ is located on a circle with radius $\sqrt{2} |\alpha|$ in the phase plane, as shown in Fig. \ref{fig:circle}.
	
	\subsection{Quantum walk for Heisenberg-Weyl group}
	
	The quantum walk over a circle on the phase plane arises naturally for a harmonic oscillator. 
	The harmonic oscillator serves as the walker whose location is determined by the HW coherent state.
	As shown in Fig. \ref{fig:circle}, a set of equally displacing points on the circle can be expressed as $\left\{\ket{\alpha_n}\right\}$, with $\alpha_n = |\alpha| \rme^{\rmi \theta_n}$, $\theta_n = n \delta \theta$, $n \in \left[-\frac{L}{2}, \frac{L}{2}\right]$ and $\delta \theta = 2 \pi / L$. $L$ is the total number of sites on the phase plane.
	
	In addition to the walker, a typical discrete quantum walk also consists of a coin which can be an arbitrary two-level system. For example, a spin can be regarded as the coin, with spin-up ($\ket{\uparrow}$) and spin-down ($\ket{\downarrow}$) states corresponding to the head and tail of the coin respectively. In each step of the quantum walk, one needs to flip the coin at first. This process can be described by the coin-flip operator $\hat{C}$. In this paper, we employ the widely-used Hadamard gate $\hat{H}$ to perform the coin-flip operation. $\hat{H}$ can be written as a $2 \times 2$ matrix in the bases of spin-up and spin-down states as follows,
	\begin{eqnarray} 
		\hat{H} =  \frac{1}{\sqrt{2}}
		\left(
		\begin{array}{cc}
			1 & 1 \\
			1 & -1
		\end{array}
		\right) .
	\end{eqnarray}
	with which the coin-flip operator can be expressed as
	\begin{eqnarray} \label{eq:cf_coherent}
		\hat{C} = \hat{I}_{\text{a}} \otimes \hat{H}.
	\end{eqnarray}
	
	Subsequently, the walker shifts its location according to the state of the flipped coin. This process is described by the conditional-shift operator $\hat{S}_\text{a} (\delta \theta)$, which can be written as
	\begin{eqnarray}
		\hat{S}_\text{a} (\delta \theta) &=& \exp \left(\rmi \delta \theta \hat{a}^{\dagger} \hat{a} \otimes \hat{\sigma}_z \right) \\
		&=& \rme^{+\rmi \delta \theta \hat{a}^{\dagger} \hat{a}} \otimes \ket{\uparrow} \bra{\uparrow} + \rme^{-\rmi \delta \theta \hat{a}^{\dagger} \hat{a}} \otimes \ket{\downarrow} \bra{\downarrow} , \nonumber
	\end{eqnarray}
	with $\hat{\sigma}_z$ the Pauli matrix. The conditional-shift operator $\hat{S}_\text{a} (\delta \theta)$ leads to a counter-clockwise rotating by angle $\delta \theta$ when the coin is on the spin-up state, and vice versa, as follows,
	\begin{subequations}
	\begin{eqnarray}
		\hat{S}_\text{a} (\delta \theta) \ket{\alpha_n} \otimes \ket{\uparrow} &=& \ket{\alpha_{n + 1}} \otimes \ket{\uparrow}, \\
		\hat{S}_\text{a} (\delta \theta) \ket{\alpha_n} \otimes \ket{\downarrow} &=& \ket{\alpha_{n - 1}} \otimes \ket{\downarrow} .
	\end{eqnarray}
	\end{subequations}
	
	Given that the walker and the coin are initially at $\ket{\alpha_0}$ and $\ket{\uparrow}$ respectively, after $l$ steps the state of the whole system will become 
	\begin{eqnarray}
		\ket{\psi(l)} =\left(\hat{S}_\text{a} (\delta \theta) \cdot \hat{C}\right)^l \ket{\psi (0)}, 
	\end{eqnarray}
	with $\ket{\psi (0)} = \ket{\alpha_0} \otimes \ket{\uparrow}$. For an idealized quantum walk, the walker's states corresponding to different sites are orthogonal. In other words, there is no overlap between different states. However, the coherent states are not orthogonal due to
	\begin{eqnarray}
		\interproduct{\alpha_m}{\alpha_n} = \exp \left[-|\alpha|^2 \left(1 - e^{i (n - m) \delta \theta}\right)\right] ,
	\end{eqnarray}
	which leads to a limitation on the number of steps for the quantum walk \cite{PhysRevA.67.042305,PhysRevLett.103.090504,Matjeschk_2012}. 
	
	\section{Quantum walk related to the SU(1,1) group} \label{sec:SU11}
	
	\begin{figure}[htb]
		\centering
		\includegraphics[scale=0.4]{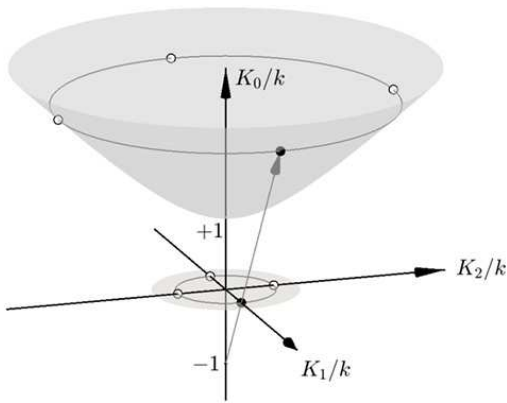} 
		\caption{Schematic diagram of the quantum walk on a circle  related to the SU(1,1) group. The walker located on the hyperboloid with $\left(K_1, K_2, K_0\right)$ is described by the SU(1,1) coherent states $\ket{k, \zeta}$. Alternately, one can choose the Poincar\'{e} disk which can be regarded as a projection viewed from $(0, 0, -1)$.}\label{fig:hyperboloid}
	\end{figure}
	
	\subsection{SU(1,1) group}
	
	The SU(1,1) group is non-compact. The generators associated with SU(1,1) group satisfy the following commutation relations,
	\begin{eqnarray}
		\left[\hat{K}_0, \hat{K}_{\pm}\right] = \pm \hat{K}_{\pm}, \qquad \left[\hat{K}_+, \hat{K}_-\right] = -2 \hat{K}_0 .
	\end{eqnarray}
	The Casimir $\hat{C}$ operator commutes with all the generators of the su(1,1) Lie algebra, which can be written as
	\begin{eqnarray}
		\hat{C} &=& \hat{K}_0^2 - \frac{1}{2} \left(\hat{K}_+ \hat{K}_- + \hat{K}_- \hat{K}_+\right) .
	\end{eqnarray}
	One can choose the basis $\ket{k,m}$, which satisfies
	\begin{subequations}
		\begin{eqnarray}
			\hat{K}_0 \ket{k,m} &=& (k + m) \ket{k,m}, \\
			\hat{K}_+ \ket{k,m} &=& \sqrt{(m + 1)(m + 2k)} \ket{k, m + 1}, \\
			\hat{K}_- \ket{k,m} &=& \sqrt{m(m + 2k - 1)} \ket{k, m - 1}, \\
			\hat{C} \ket{k,m} &=& k (k - 1) \ket{k, m} ,
		\end{eqnarray}
	\end{subequations}
	with $m = 0, 1, 2, \dots$ The number $k$ is known as the Bargmann index which separates different irreducible representations. 
	
	The HW coherent state defined in the preceding section can be generalized to any Lie group.	The SU(1,1) coherent state is defined as  \cite{Perelomov1986}
	\begin{eqnarray} \label{eq:su11_CS}
		\ket{k, \zeta} &=& \exp \left(\zeta^* \hat{K}_+ - \zeta \hat{K}_-\right) \ket{k, 0} ,
	\end{eqnarray}
	with $\zeta = r \rme^{\rmi \theta}$. It's easy to prove that 
	\begin{equation}
		\left(K_0, K_1, K_2\right) = k \left(\cosh 2r, \sinh 2r \cos \theta, \sinh 2r \sin \theta\right),
	\end{equation}
	with $K_i = \bra{k, \zeta} \hat{K}_i \ket{k, \zeta}$ ($i=0,1,2$) and
	\begin{eqnarray}
		\hat{K}_1 = \frac{\hat{K}_+ + \hat{K}_-}{2}, \quad \hat{K}_2 = \frac{\hat{K}_+ - \hat{K}_-}{2 \rmi} .
	\end{eqnarray}
	Therefore, $K_0^2  - K_1^2 - K_2^2 = k^2$ indicates that $\ket{k, \zeta}$ is centered at $\left(K_1, K_2, K_0\right)$ on the hyperboloid, as shown in Fig. \ref{fig:hyperboloid}. Alternately, one can choose the Poincar\'{e} disk to geometrize each SU(1,1) coherent state. As long as  $r$ doesn't change, $K_0$ is fixed whereas $K_1^2 + K_2^2 = k^2 \sinh^2 2r$ forms a circle on the hyperboloid or the Poincar\'{e} disk. A set of equally displacing sites on the circle can be expressed as $\left\{\ket{k, \zeta_n}\right\}$, with $\zeta_n = r e^{i \theta_n}$.
	
	\subsection{Quantum walk for SU(1,1)}
	
	To perform the quantum walk over a circle on the hyperboloid, one can choose the same coin-flip operator as Eq. (\ref{eq:cf_coherent}). With regard to the conditional-shift operator $\hat{S}  (\delta \theta)$, it can be expressed as
	\begin{eqnarray}
		\hat{S}  (\delta \theta) &=& \exp \left(-\rmi \delta \theta \hat{K}_0 \otimes \sigma_z \right) \\
		&=& e^{-\rmi \delta \theta \hat{K}_0} \otimes \ket{\uparrow} \bra{\uparrow} + e^{+\rmi \delta \theta \hat{K}_0} \otimes \ket{\downarrow} \bra{\downarrow} , \nonumber
	\end{eqnarray}
	which leads to
	\begin{subequations}
		\begin{eqnarray}
			\hat{S} (\delta \theta) \ket{k, \zeta_n} \otimes \ket{\uparrow} &=& \ket{k, \zeta_{n + 1}} \otimes \ket{\uparrow}, \\
			\hat{S} (\delta \theta) \ket{k, \zeta_n} \otimes \ket{\downarrow} &=& \ket{k, \zeta_{n - 1}} \otimes \ket{\downarrow} .
		\end{eqnarray}
	\end{subequations}
	It should be noted that the SU(1,1) coherent states are also not orthogonal due to
	\begin{eqnarray} \label{eq:su11_nonorth}
		\interproduct{k, \zeta_m}{k, \zeta_n} = \left[\cosh^2 r - e^{i \left(\theta_m - \theta_n\right)} \sinh^2 r \right]^{-2k} .
	\end{eqnarray}
	However, we can decrease the overlap $\left|\interproduct{k, \zeta_m}{k, \zeta_n}\right|$ by increasing $k$ and $r$. 
	
	\subsubsection{One-mode realization}
	
	Depending on the underlying physical system, we can choose different realizations of the SU(1,1) group. The generators of su(1,1) Lie algebra with one-mode realization \cite{Perelomov1986} can be written as
	\begin{subequations}\label{eq:one-mode}
		\begin{eqnarray}
			\hat{K}_0 &=& \frac{1}{2} \left(\hat{a}^{\dagger} \hat{a} + \frac{1}{2}\right), \\
			\hat{K}_+ &=& \frac{1}{2} \left(\hat{a}^{\dagger}\right)^2, \\
			\hat{K}_- &=& \frac{1}{2} \hat{a}^2 .
		\end{eqnarray}
	\end{subequations}
	The corresponding Bargmann index is $k = \frac{1}{4}$ or $\frac{3}{4}$. Given the Fock states $\ket{n}_{\text{a}}$ ($\hat{a}^{\dagger} \hat{a} \ket{n}_{\text{a}} = n \ket{n}_{\text{a}}$ with $n=0,1,2\dots$), $\ket{k, m}$ can be rewritten as
	\begin{eqnarray}
		\ket{k, m} = \ket{2 \left(m + k - \frac{1}{4}\right) }_{\text{a}}. 
	\end{eqnarray}
	Accordingly, the SU(1,1) coherent state is nothing but the squeezed vacuum state for $k = \frac{1}{4}$ and squeezed one-photon state for $k = \frac{3}{4}$, due to
	\begin{eqnarray} \label{eq:one-mode_CS}
		\ket{k, \zeta} &=& \exp \left[ \frac{\zeta^*}{2} \left(\hat{a}^{\dagger}\right)^2 - \frac{\zeta}{2} \hat{a}^2 \right] \ket{2 \left(k - \frac{1}{4}\right) }_{\text{a}} .
	\end{eqnarray}
	Both of them can be produced in the experiments by engineering the interactions between a quantum system (such as a trapped ion) and the environment \cite{doi:10.1126/science.1261033,PhysRevLett.119.033602}.
	It should be noted that the SU(1,1) coherent state [Eq. \ref{eq:one-mode_CS}] is always centered at the origin of the phase plane, namely, $\left(x, p\right) = (0,0)$. Therefore, the circular movement of walker cannot be visualized on the phase plane but rather on the hyperboloid.
	
	In the one-mode realization, the conditional-shift operator $\hat{S}_{1} (\delta \theta)$ can be written as
	\begin{eqnarray}
		\hat{S}_{1} (\delta \theta) &=& \exp \left(-\rmi \delta \theta \hat{K}_0 \otimes \sigma_z \right) \\
		&=& \hat{S}_{\text{a}} \left(-\frac{\delta \theta}{2}\right) \cdot \left[ \hat{I}_{\text{a}} \otimes \exp\left(-\rmi \frac{\delta \theta}{4} \hat{\sigma}_z\right)\right]. \nonumber
	\end{eqnarray}
	Therefore, the conditional-shift operator $\hat{S}_1 (\delta \theta)$ in the one-mode realization is equivalent to the coin operation $ \hat{I}_{\text{a}} \otimes \exp \left(-\rmi \delta \theta \hat{\sigma}_z / 4\right)$ followed by the conditional-shift operator $\hat{S}_{\text{a}} \left(-\frac{\delta \theta}{2}\right)$ for the HW coherent state. 
	
	In a word, the quantum walk for Heisenberg-Weyl group can be easily extended to that for SU(1,1) in the one-mode realization. The latter only introduces an extra coin operation before performing the conditional shift for the walker. Besides, the squeezed vacuum or one-photon state should be introduced as the initial state.
	
	\subsubsection{Two-mode realization}
	The generators of su(1,1) Lie algebra with two-mode realization \cite{Perelomov1986} can be expressed as
	\begin{subequations} \label{eq:two-mode}
	\begin{eqnarray} 
		\hat{K}_0 &=& \frac{1}{2} \left(\hat{a}^{\dagger} \hat{a} + \hat{b}^{\dagger} \hat{b} + 1\right), \\
		\hat{K}_+ &=& \hat{a}^{\dagger} \hat{b}^{\dagger}, \\
		\hat{K}_- &=& \hat{a} \hat{b} .
	\end{eqnarray}
	\end{subequations}
	Given the Fock states $\ket{n}_{s}$ ($s=a, b$), $\ket{k, m}$ can be rewritten as
	\begin{eqnarray}
		\ket{k, m} = \ket{m + 2 k - 1}_{\text{a}} \otimes \ket{m}_{\text{b}} ,
	\end{eqnarray}
	with the Bargmann index $k = \frac{1}{2},~1,~\frac{3}{2},~\dots $
	From Eqs. (\ref{eq:su11_CS}) and (\ref{eq:two-mode}), the corresponding SU(1,1) coherent state is
	\begin{eqnarray}
		\ket{k, \zeta} = \exp \left(\zeta^* \hat{a}^{\dagger} \hat{b}^{\dagger} - \zeta \hat{a} \hat{b} \right) \ket{2 k - 1}_{\text{a}} \otimes \ket{0}_{\text{b}} ,
	\end{eqnarray}
	which represents the two-mode squeezed operator acting on the Fock states.
	
	In the two-mode realization, the conditional-shift operator $\hat{S}_{2} (\delta \theta)$ can be written as
	\begin{eqnarray}
		\hat{S}_{2} (\delta \theta) &=& \exp \left(-\rmi \delta \theta \hat{K}_0 \otimes \sigma_z \right) \\
		&=& \hat{S}_{\text{a}} \left(-\frac{\delta \theta}{2}\right) \cdot \hat{S}_{\text{b}} \left(-\frac{\delta \theta}{2}\right) \nonumber\\
		&&\cdot\left[ \hat{I}_{\text{a}} \otimes  \hat{I}_{\text{b}} \otimes \exp\left(-\rmi \frac{\delta \theta}{2} \hat{\sigma}_z\right)\right], \nonumber
	\end{eqnarray}
	which corresponds to a coin operation followed by the conditional-shift operators for mode a and b respectively.
	
	
	\subsection{Probability distribution and standard deviation}
	
	The quantum walk is a quantum mechanical extension of the classical random walk. Due to the quantum interference effect, the quantum walk presents a ballistic spread quadratically faster than the classical one which yields a diffusive spread. The quadratic enhancement can be well captured by the probability distribution and the corresponding standard deviation. The probability to find the walker at site $n$ after $l$ steps \cite{PhysRevLett.103.090504,Matjeschk_2012} can be approximately given by
	\begin{eqnarray}
		P_n (l) = \frac{1}{\mathcal{N}} \bra{k, \zeta_n} \hat{\rho}_{\text{w}} (l) \ket{k, \zeta_n} ,
	\end{eqnarray}
	where $\hat{\rho}_{\text{w}}(l) = \tr{\ket{\psi (l)} \bra{\psi (l)}}{\text{c}}$ is the reduced density matrix for the walker and $\mathcal{N}$ is introduced to normalize the whole probabilities. The approximation is due to the nonorthogonality of the SU(1,1) coherent states, as given in Eq. (\ref{eq:su11_nonorth}).

	\begin{figure}[htb]
		\centering
		\includegraphics[scale=1]{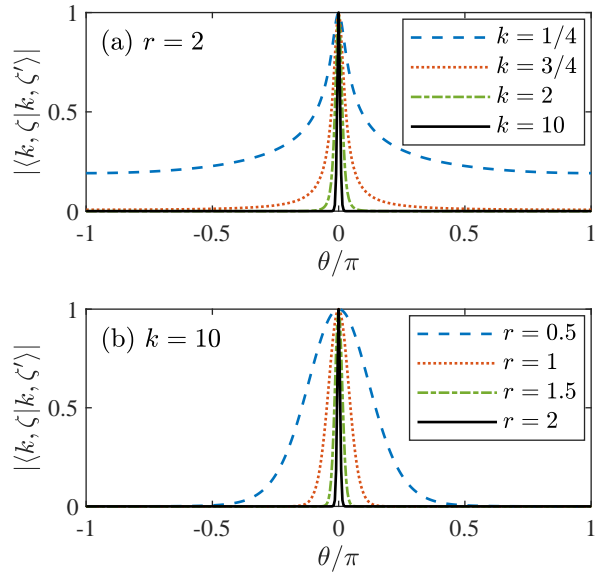} 
		\caption{The overlap of SU(1,1) coherent states $\left|\interproduct{k,\zeta}{k,\zeta'}\right|$ with $\zeta = r \rme^{\rmi \theta}$ and  $\zeta' = r$ as a function of $\theta$. (a) $r=2$ is fixed with $k = \frac{1}{4}$ (blue dash line), $\frac{3}{4}$ (red dot line), $2$ (green dash-dot line) and $10$ (black solid line). (b) $k=10$ is fixed with $r = 0.5$ (blue dash line), $1$ (red dot line), $1.5$ (green dash-dot line) and $2$ (black solid line).}\label{fig:overlap}
	\end{figure}
	
	The overlap between two SU(1,1) coherent states $\ket{k, \zeta = r \rme^{\rmi \theta}}$ and  $\ket{k, \zeta' = r}$ is depicted in Fig. \ref{fig:overlap}. When $r$ remains unchanged, the overlap decreases with the increase of $k$ as shown  in Fig. \ref{fig:overlap}(a). For the one-mode realization, the Bargmann index $k$ only has two choices. The overlap for $k = \frac{3}{4}$ is much smaller than that for $k = \frac{1}{4}$. With respect to the two-mode realization, $k$ can be much larger which further decreases the overlap. Fig. \ref{fig:overlap}(b) illustrates the influence of $r$ on the overlap. Clearly, we prefer a larger $r$ to eliminate the nonorthogonality.
	
	\begin{figure}[htb]
		\centering
		\includegraphics[scale=1]{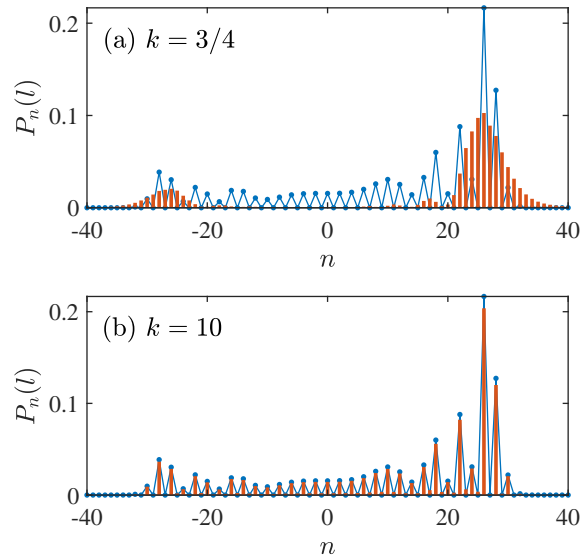} 
		\caption{Probability distribution of the quantum walk with $L=200$ sites after $l=40$ steps for $r = 2$, (a) $k = \frac{3}{4}$ and (b) $k = 10$. The red bars correspond to the quantum walk with SU(1,1) coherent states, whereas the blue dots correspond to the idealized quantum walk which serve as a benchmark.}\label{fig:probability}
	\end{figure}
	
	As the overlap decreases with the increase of the Bargmann index $k$, we expect that a larger $k$ can better simulate the idealized quantum walk. When $k$ is large enough, the overlap between different sites can be ignored, namely $\interproduct{k, \zeta_m}{k, \zeta_n} \approx \delta_{m, n}$, which will lead to the idealized quantum walk. The probability distribution after $l = 40$ steps is shown in Fig. \ref{fig:probability}.  The idealized quantum walk is also present for comparison. Initially, the walker is located at site $n=0$. For an idealized quantum walk, the walker can only be detected at sites with even position index $n$ if the number of steps $l$ is even. This constraint is relieved due to the nonorthogonality, which is especially obvious for the one-mode realization with $k = \frac{1}{4}$ or $\frac{3}{4}$. The multipeak structures are smeared out as shown in Fig. \ref{fig:probability}(a), and similar phenomena have also been found for the quantum walk with  coherent states on the phase plane \cite{Matjeschk_2012} and Bloch sphere \cite{PhysRevA.105.042215}. Beyond the one-mode realization, we can further increase $k$ in the two-mode realization. When $k$ is large enough, the quantum walk with SU(1,1) coherent states can almost exactly reproduce the idealized quantum walk over a circle, as indicated in Fig. \ref{fig:probability}(b).
	
	\begin{figure}[htb]
		\centering
		\includegraphics[scale=1]{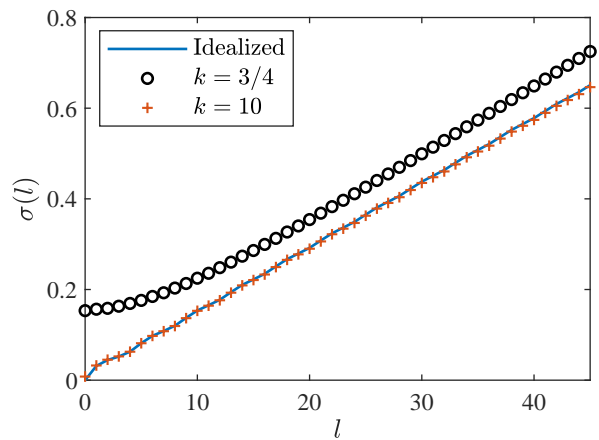} 
		\caption{The standard deviation $\sigma(l)$ as a function of number of steps $l$ for $k = \frac{3}{4}$ (black circle) and $k = 10$ (red crossing). The blue solid line corresponds to the idealized quantum walk which serve as a benchmark.}\label{fig:SD}
	\end{figure}

	Once the probability distribution is achieved, it is straightforward to calculate the standard deviation as follows,
	\begin{eqnarray}
		\sigma (l) = \sqrt{\expect{\theta^2} - \expect{\theta}^2},
	\end{eqnarray}
	with $\expect{\theta^m} = \sum_n P_n (l) \theta_n^m$. As shown in Fig. \ref{fig:SD}, the standard deviation grows linearly with the number of walking steps ($\sigma \propto l$)  which is a typical character of the quantum walk, even if the overlap is not negligible for $k = \frac{3}{4}$. 
	
	\section{Quantum Entanglement} \label{sec:entanglement}
	
	The quantum entanglement is a key ingredient in quantum information and quantum computation. Entanglement between the coin and the walker is one of the unique characters owned by the discrete quantum walk, which cannot be exactly reproduced by its classical counterpart \cite{Venegas-Andraca2012,Carneiro_2005,PhysRevA.73.042302}.
	
	The von Neumann entropy, also known as the entropy of entanglement, can be employed to quantify the entanglement between the coin and the walker \cite{Carneiro_2005,PhysRevA.73.042302}. Generally, one needs to calculate the reduced density matrix of the coin or the walker to obtain the von Neumann entropy. Since the coin is a two-level system, we can employ an alternate approach \cite{PhysRevA.68.034301}, defined as
	\begin{eqnarray}
		\mathcal{S}_E = -p_+ \log_2 p_+ - p_- \log_2 p_-,
	\end{eqnarray}
	with
	\begin{eqnarray}
		p_{\pm} = \frac{1}{2} \left(1 \pm \sqrt{M_x^2 + M_y^2 + M_z^2}\right),
	\end{eqnarray}
	and $M_{i} = \bra{\psi(l)} \hat{\sigma}_i \ket{\psi(l)}$ ($i=x,y,z$). For the Hadamard gate we considered, Carneiro et al. numerically studied the long-time asymptotic behavior of entanglement, and they conjectured that the entanglement is 0.872 
	for arbitrary coin initial states \cite{Carneiro_2005}.  Soon after that, it was proved by Abal et al. through an analytically approach in the Fourier representation \cite{PhysRevA.73.042302}.

	\begin{figure}[htb]
		\centering
		\includegraphics[scale=1]{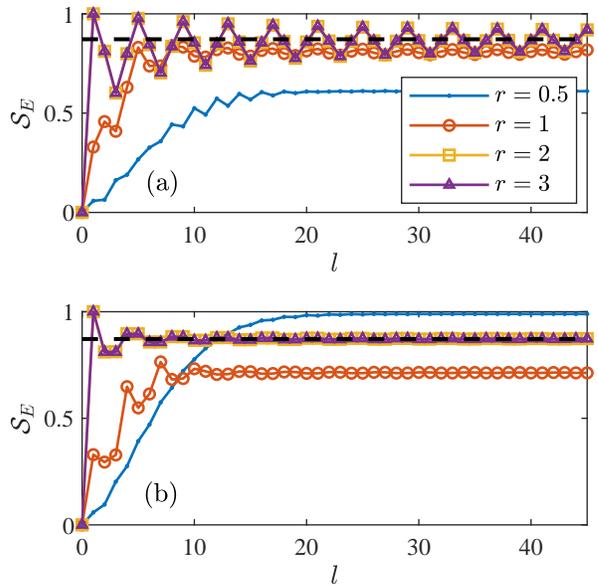} 
		\caption{The evolution of the von Neumann entropy with initial states (a) $\ket{\psi(0)}= \ket{k, \zeta_0} \otimes \ket{\uparrow}$ and (b) $\ket{\psi(0)} = \ket{k, \zeta_0} \otimes \frac{1}{\sqrt{2}} \left(\ket{\uparrow} + \rmi \ket{\downarrow}\right)$. $k=10$ is fixed with $r=0.5$ (blue dot), $1$ (red circle), $2$ (yellow square) and $3$ (purple triangle). The black dashed lines correspond to the asymptotic value $\mathcal{S}_E=0.872$ of the idealized quantum walk which serve as a benchmark.}\label{fig:entanglement}
	\end{figure}
	The evolution of the von Neumann entropy $\mathcal{S}_E$ with initial states $\ket{\psi(0)}= \ket{k, \zeta_0} \otimes \ket{\uparrow}$ and $\ket{\psi(0)} = \ket{k, \zeta_0} \otimes \frac{1}{\sqrt{2}} \left(\ket{\uparrow} + \rmi \ket{\downarrow}\right)$ is shown in Fig. \ref{fig:entanglement}  (a) and (b) respectively. When $r$ is large enough, the long-time asymptotic value tend to be $0.872$ in both cases which doesn't depend on the initial states of the coin. When $r$ is small, however, the initial states play a significant role on the long-time evolution. 
	As illustrated by Abal et al \cite{PhysRevA.73.042302}, the asymptotic entanglement depends on the initial condition of the coin when nonlocal initial conditions of the walker are considered. It should be noted that the walker is not absolutely localized due to the nonorthogonality $\interproduct{k,\zeta_0}{k, \zeta_n} \ne 0$ especially for small $r$, which leads to the dependence of the asymptotic entanglement on the initial conditions of the coin. 
	
	\section{Conclusions} \label{sec:summary}
	
	The quantum walk can be visualized by choosing proper phase spaces. Persistent attention has been drawn to the quantum walk over a circle on the phase plane, which associates with the Heisenberg-Weyl group. In this paper, we propose a new possibility to perform the quantum walk in the experiments, which generalize it to the SU(1,1) group. The walker is described by the SU(1,1) coherent states, whose movement during the walking process can be visualized on the hyperboloid or the Poincar\'{e} disk.
	
	We consider two different realizations of the SU(1,1) group. In the one-mode realization, the conditional-shift operators for the Heisenberg-Weyl and the SU(1,1) groups share similar structures. One only needs to be careful about the initial state of the walker, as the SU(1,1) coherent states correspond to the squeezed vacuum or the one-photon states. In the two-mode realization, the conditional-shift operations for both modes should be performed separately. 
	
	The nonorthogonality of the SU(1,1) coherent states precludes the quantum walk for SU(1,1) from idealized quantum walk. Increasing the Bargmann index $k$ can decrease the overlap between different SU(1,1) coherent states. From this perspective, the two-mode realization is able to provide a larger Bargmann index $k$ required to simulate the idealized quantum walk. 
	
	The quantum walk is known for its ballistic spread, which is confirmed by the quadratically growing variance with the number of steps, namely, $\sigma^2 \propto l^2$. The entanglement between the coin and the walker is also a characteristic feature of the discrete quantum walk, whose long-time asymptotic behavior depends on the coin's initial state if the nonorthogonality of the SU(1,1) coherent states is not negligible. The decoherence can destroy the quantum interference and entanglement, which leads to the classical random walk with $\sigma^2 \propto l$ \cite{PhysRevA.67.032304,PhysRevLett.91.130602}. The dependence of the quantum walk for SU(1,1) on the decoherence deserves further studies, which is left to future research.
	
	\begin{acknowledgments}
		This research was supported by Zhejiang Provincial Natural Science Foundation of China under Grant No. LQ23A050003.
	\end{acknowledgments}
	
	%
	
\end{document}